\documentclass[pra,aps,twocolumn,superscriptaddress,showpacs]{revtex4-1}

\usepackage[bookmarksnumbered, colorlinks=true, allcolors=blue, plainpages]{hyperref}
\usepackage{graphicx}
\usepackage{xcolor}
\usepackage{amssymb}
\usepackage{bm}
\usepackage{amsmath}
\usepackage{epstopdf}
\usepackage{revsymb}
\usepackage{mathptmx}
\DeclareMathAlphabet{\mathcal}{OMS}{cmsy}{m}{n}

\newcommand{\tr}[0]{\text{Tr}}
\newcommand{\mean}[1]{\langle#1\rangle}
\newcommand{\ket}[1]{|#1\rangle}

\begin{document}
\title{Non-Gaussian macroscopic entanglement of motion in a hybrid electromechanical device}
\author{Najmeh Es'haqi-Sani}
\email{na.eshaqisani@mail.um.ac.ir}
\affiliation{Department of Physics, Ferdowsi University of Mashhad, Mashhad, PO Box 91775-1436, Iran}
\author{Mehdi Khazaei Nezhad}
\email{khazaeinezhad@um.ac.ir}
\affiliation{Department of Physics, Ferdowsi University of Mashhad, Mashhad, PO Box 91775-1436, Iran}
\author{Mehdi Abdi}
\email{mehabdi@cc.iut.ac.ir}
\affiliation{Department of Physics, Isfahan University of Technology, Isfahan 84156-83111, Iran}

%\date{\today}

\begin{abstract}
We propose a scheme for entangling the motion of two massive objects in a hybrid electromechanical architecture.
The entanglement is generated due to the interaction of two mechanical oscillators with a mediating superconducting qubit.
We show that the generated macroscopic entangled states are non-Gaussian and its lifetime is limited by coherence time of the qubit.
The entanglement is attainable in a wide range of parameters with appropriate control of the qubit.
We confirm performance of our scheme by numerically solving the quantum optical master equation including sources of noise.
The effect of imperfections, such as asymmetries in the coupling rates as well as mechanical thermal noise are studied and shown how they affect the amount and lifetime of the entangled state.
Due to the nonlinear nature of the qubit, the initial Gaussian state of mechanical resonators evolves into a quasi-stationary non-Gaussian state, which is essential for universal quantum information processing in continuous variable systems.
This work, therefore, provides the first step towards a universal continuous variable quantum network.
\end{abstract}

%\textcolor{green}{text}
%\pacs{???}

\maketitle

\section{Introduction}
Non-classical states such as entangled states in macroscopic scales are very fragile due to the environmental disturbances and their creation counts as an outstanding task as they are beneficial for exploration of the quantum to classical boundary as well as quantum information processing~\cite{KCLee, Manciniprl}.
Generating nonclassical quantum states of motion, especially creation of quantum entanglement has attracted much attention in the recent years~\cite{Houhou2015, Li2015, Wang2016, Asadian2016, Abdi2016}.
The quantum information theory exploits it as an important physical resource to carry out numerous quantum computational and communication tasks~\cite{Wootters}.

Generation and stabilization of entanglement in massive objects like mechanical resonators (MRs) is very laborious because of their rapid decoherence induced by the environment which is hardly controllable in large-scale systems.
At the mesoscopic level, the entanglement has been realized in different systems such as two atomic ensembles \cite{Julsgaard:atom}, an electromechanical architecture \cite{Palomaki:2013electrome}, and in two Josephson-junction qubits \cite{Steffen;J,Berkley;J}. The Gaussian entanglement at the macroscopic level between two MRs has been investigated theoretically in optomechanical setups \cite{J.Li et al.;2015,J.Li;2017} and also recently, an experimental demonstration of generation and stabilization of such entangled states has been reported~\cite{OckeloenKorppi;nature;2018}.

In the above mentioned works, the entangled states of motion are Gaussian. That is, their characteristic and quasiprobability distribution functions in the phase space are Gaussian~\cite{Adesso}. In general, Gaussian states can be experimentally prepared with a high degree of control especially in quantum optical setups. In spite of belonging to an infinite dimensional Hilbert space, their properties are easily handled in theoretical studies as they are completely described by the first and second moments of their canonical operators~\cite{Weedbrook;Gaussian}.
Although they are counted as a useful resource for continuous variable (CV) quantum information processing, there are several tasks that demand employing non-Gaussian states and/or operations. In fact, no-go theorems prohibit CV entanglement distillation~\cite{Fiur;D,Eisert} and it has been proven that non-Gaussianity is essential for universal CV quantum computation~\cite{Menicucci}.
Moreover, it is proven that it can help to improve the efficiency of other quantum information tasks such as quantum teleportation~\cite{Opatrny;QT,DEll'Anno}, security~\cite{Lee2019}, cloning \cite{Cerf;cloning}, and to test quantum nonlocality by the violation of Bell's inequality~\cite{Abouraddy;BI}.
Notably, compared to their Gaussian counterparts, the non-Gaussian entangled states are robuster against environmental effects~\cite{Kumar et al.;2011}.
Macroscopic non-Gaussian states are useful for force sensing~\cite{Caves} and capturing signatures of gravitational effects on quantum systems~\cite{Bose2017}.

The interplay of various types of interactions in hybrid quantum systems have provided the possibility of preparing various non-classical states in different components of system~\cite{Genes, Montenegro,Mintert,Kounalakis}.
Novel strategies based on hybrid systems have also been proposed for quantum non-demolation measurement of MRs~\cite{Viennot} as well as achieving strong and tunable coupling regimes for their quantum control~\cite{Rodrigues,Kounalakis}.
Despite few proposed schemes, generation of the \textit{macroscopic non-Gaussian entangled states of motion} has remained widely overlooked.
Generally, setups in which motion is coupled to a nonlinear quantum object, such as superconducting qubits, can open up the possibility of generating, manipulating, and storing non-Gaussian states in mechanical degrees of freedom~\cite{Chu, Abdi2017}. Here, we propose a device based on our previous work~\cite{Abdi et al.PRL;2015} to generate non-Gaussian entangled states of two MRs in a superconducting circuit.
Our hybrid device is composed of a superconducting transmon qubit coupled to two nanobeams in its shunt capacitance. The qubit is driven via a strongly coupled superconducting coplanar waveguide resonator.
Superconducting circuits are of interest in implementing quantum information processing due to their low intrinsic dissipation and their nonlinear nature~\cite{Armour;2002,Nigg;2012}. Transmon qubits are charge-insensitive superconducting qubits with sufficient anharmonicity for selective qubit control~\cite{Koch}. Thus, very appropriate for hybridization with MRs.
We show that one can overcome the difficulty of creating the non-Gaussian nonclassical states in a linear resonator by interposing the nonlinearity of superconducting qubits and appropriately driving the qubit.
We analytically elaborate on the possibility of creating entangled states of MRs via the qubit.
The numerical results obtained by solving the quantum optical master equation verify generation of the two-mode entangled non-Gaussian state of the motion in the macroscopic scale. The lifetime of which dependents on: First, the amount of asymmetry in their coupling strengths to the transmon qubit. Second, the coherence time of the intermediating qubit. Third, the thermal noise of mechanical resonators.
Our simulations suggest that the highest entanglement is attained for a fully symmetric system. Moreover, the longer the coherence time of the transmon qubit, the longer the MRs remain entangled. Finally, the thermal noise affecting the MRs should be controlled by a cooling mechanism, e.g. sideband cooling~\cite{Rabl2010}.
Also, we show that the quasi-stationary state of the two-mode mechanical system has a finite non-Gaussianity.  
 
The outline of the paper is as follows. In section~\ref{model}, the  general model and the system Hamiltonian is introduced. In section~\ref{fmec}, we obtain the effective Hamiltonian describing the fully mechanical subsystem with eliminating the degrees of freedom of the transmon mode through applying canonical Fr{\"o}hlich-Nakajima transformation. We explain the measure used to quantify the non-Gaussianity of the fully mechanical state as well as its entanglement in Sec.~\ref{measures}. Section~\ref{nr} is devoted to numerical results related to the system under survey in the case of an experimentally feasible scenario, and concluding remarks is given in section~\ref{conclusion}.

%%%%%%%%%%%%%%%%%%%%%%%%%%%%%%%%%
\section{The Model \label{model}}

\begin{figure}[t!]
\includegraphics[angle=0, width=1.0\linewidth]{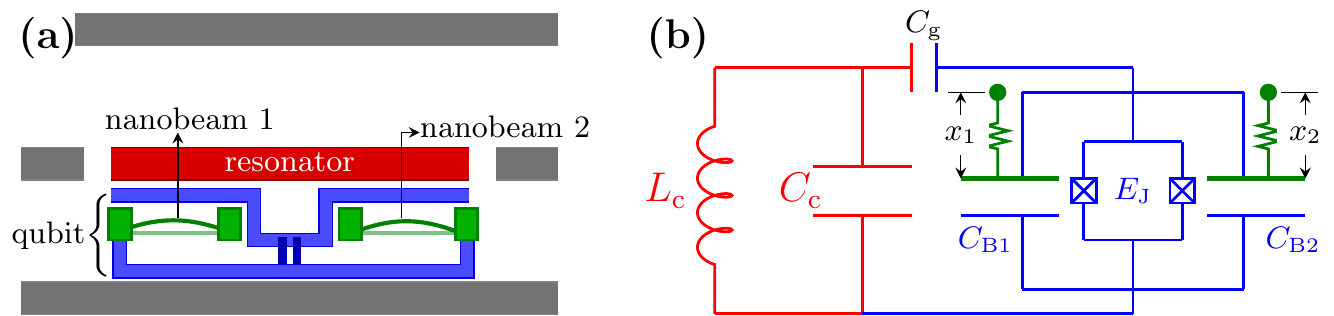}

\caption{(color online)
(a) Scheme of the hybrid electromechanical device: a superconducting transmon qubit with two mechanical resonators at its shunt capacitance, and superconducting coplanar transmission line resonator. (b) The equivalent circuit diagram.}
\label{fig1}
\end{figure}

We consider an electromechanical hybrid system composed of a superconducting coplanar transmission line resonator as an $L_{c}C_{c}$ oscillator capacitively coupled to a transmon qubit by the gate capacitance $C_{g}$. This system has been illustrated schematically in Fig.~\ref{fig1}. The superconducting transmon qubit of the proposed setup consists of a shunt capacitance decomposed into $C_{B1}$ and $C_{B2}$ that both have a part free to oscillate. As a result, the capacitance energy of the qubit depends on position of the two mechanical resonators: MR1 and MR2. This couples the MRs to the qubit and the transmon qubit mediates an interaction between two resonators [see the Appendix for more details].
The system Hamiltonian in rotating wave approximation (RWA) is given by ($\hbar=1$)~\cite{Abdi et al.PRL;2015}:
\begin{subequations}
\begin{eqnarray}
\label{eq1}
\hat{H}&=&\hat{H}_{0}+\hat{H}_{1}+\hat{H}_{d}, \\
\hat{H}_{0}&=&\omega_{t}\hat{a}^{\dagger}\hat{a} -\lambda \hat{a}^{\dagger 2}\hat{a}^{2}+\omega_{1}\hat{b}_{1}^{\dagger}\hat{b}_{1}+\omega_{2}\hat{b}_{2}^{\dagger}\hat{b}_{2}, \\
\hat{H}_{1}&=&g_{01}\hat{a}^{\dagger}\hat{a}(\hat{b}_{1}+\hat{b}_{1}^{\dagger})+g_{02}\hat{a}^{\dagger}\hat{a}(\hat{b}_{2}+\hat{b}_{2}^{\dagger}), \\
\hat{H}_{d}&=&\big[E_{1}(t)+E_{2}(t)\big]\hat{a}^{\dagger}e^{-i\omega_t t}+ \text{H.c.},
\end{eqnarray}
\end{subequations}
where $\omega_{t}=E_{C}(\sqrt{8\zeta}-1)$ (with $\zeta=E_{J}/E_{C}$) is the transmon transition frequency between the ground and its first excited states. Here, the superconducting transmon qubit is modeled as a Duffing anharmonic oscillator with negative nonlinearity $\lambda=E_{C}/2$ and annihilation (creation) operator $\hat{a}$($\hat{a}^{\dagger}$) and the RWA is valid for $\lambda \ll \omega_t$. $E_{J}$ and $E_{C}=e^2/2C_{\varSigma}$ are the Josephson and charging energies of the qubit, respectively. Here, $C_{\varSigma}=C_{g}+C_{B1}+C_{B2}+C_{J}$ with $C_{J}$ the capacitance of the Josephson junction. The annihilation (creation) operator of the MR$j$ is $\hat{b}_{j}$($\hat{b}_{j}^{\dagger}$) with the corresponding displacement operator $\hat{x}_{j}=x_{{\rm zp},j}(\hat{b}_{j}+\hat{b}_{j}^{\dagger})$ with $j=1,2$, where $x_{{\rm zp},j}=\sqrt{\hbar/2m_{j}\omega_{j}}$ is the zero-point amplitude of oscillator $j$ with the effective mass $m_{j}$. 
The qubit--mechanical coupling strengths are introduced by $g_{0j}\equiv\sqrt{2\zeta} E_{C} (C_{Bj}/C_{\Sigma})(x_{{\rm zp},j}/d_{0j})$ with $d_{0j}$ the equilibrium distance between the nanobeams and plates of the shunt capacitors.

In order to entangle the two resonators, we drive the transmon mode bichromatically with two time-dependent amplitudes $E_{1}(t)$ and $E_{2}(t)$. As it will become clear later, the frequency of these two drives $E_{1}(t)$ and $E_{2}(t)$ are chosen such that the two-mode squeezing (TMS) becomes the resonant process.
In Sec.~\ref{nr} it is shown by the numerical calculations that at the considered parameter region the long enough coherence time for transmon qubit facilitates the achievement of entanglement of motion of the mechanical resonators. In the operation regime of our setup $E_{J}/E_{C} \gg 1$ and also is $g_{01},g_{02}\ll \omega_{t}$, therefore, the RWA is valid.
%%%%----------figure1--------------
%%%%-------------------------------

Since the considered setup is generally an open quantum system, its full dynamics is only correctly explained when the dissipative incoherent processes induced by environment such as relaxation and dephasing are included. The transmon is subject to the energy relaxation that occurs with the rate $\gamma_{t}=1/T_{1}$. Its total decoherence rate is given by $1/T_{2}^{*}=1/(2T_{1})+1/T_{\phi}$ in which $\gamma_{\phi}=1/T_{\phi}$ is the pure dephasing rate. Currently, high quality superconducting qubits with relaxation and dephasing times about ${T_{1}} \approx~50~\mu$s and ${T_{2}}^{*} \approx~20~\mu$s are realistic and even with housing the transmon qubit inside a 3D superconducting cavity \cite{Paik}, energy relaxation times $T_{1}> 100~\mu$s have been observed \cite{Riste,Dial,Narla}. Also, in Ref.~\cite{ZWang} an effective method for protecting transmon qubit against photon noise has been suggested.

Meanwhile, the MRs are coupled to their respective thermal bath with damping rates $\gamma_{jm}=\omega_{j}/Q_{j}$. Here, $Q_{j}$ is the mechanical quality factor of $j$th resonator.
The mechanical nanobeam resonators with the fundamental vibrational mode frequencies in the range of $1~\text{MHz}-1~\text{GHz}$ with the quality factors as high as $Q = 10^{6}$ and more have been realized~\cite{Rodrigues}.

The full dynamics of our system is described by the following quantum optical master equation:
\begin{eqnarray}
\label{eq2}
\dot\rho &=& -i[\hat{H},\rho]+\gamma_{t}D_{\hat{a}}\rho+\gamma_{\phi}D_{\hat{a}^{\dagger}\hat{a}}\rho
\nonumber\\
&&+\sum_{j=1,2}(\bar n_{j}+1)\gamma_{j}D_{\hat{b}_{j}}\rho +\bar n_{j}\gamma_{j}D_{\hat{b}_{j}^{\dagger}}\rho~~,
\end{eqnarray}
in which the superoperator $D_{\hat{A}}[\rho]=\hat{A}\rho\hat{A}^{\dagger}-\frac{1}{2}\{\hat{A}^{\dagger}\hat{A},\rho\}$ describes dissipation of a general system operator $\hat{A}$ and the thermal occupation number of resonator $j$ is given by $\bar n_{j}=\big[\exp(\hbar\omega_{j}/k_{\rm B}T)-1\big]^{-1}$ that $k_{\rm B}$ and $T$ are the Boltzmann constant and the temperature of the thermal bath coupled to mechanical resonators, respectively. In this work, we assume very low (about a few mK) working temperatures such that the transmon qubit remains in its ground state.
By solving the master equation (\ref{eq2}), the density matrix describing of the whole system at every instance of time $\rho(t)$ is attained. The state of fully mechanical bipartite subsystem, $\rho_{12}(t)$, is then obtained by tracing over the qubit part. One, intuitively expects to get a non-Gaussian state for the harmonic MRs because of their coupling to a source of quantum nonlinearity, the transmon qubit. This is indeed confirmed to be the case in our numerical simulations.

%%%%%%%%%%%%%%%%%%%%%%%%%%%%%%%%%
\section {EFFECTIVE HAMILTONIAN \label{fmec}}

In order to obtain a clearer picture of the mechanical--mechanical interaction mediated by the transmon qubit, we eliminate the degree of freedom of the transmon mode $\hat{a}$ via a polaron transform.
For this purpose, one uses the Fr{\"o}hlich-Nakajima transformation which is a unitary transformation widely used in the condensed matter physics~\cite{Frohlich,Nakajima} and quantum optics~\cite{Wang;2015,Sun;2006,Y.Li;2007}.
In this part, for the sake of clarity and simplicity, we approximate the transmon qubit with a two-level-system by only considering its ground and first excited states. Nevertheless, the full Hamiltonian will be recalled in the numerical simulations.

We first assume a general form for the amplitudes of bichromatic drives applied to the transmon mode. We then will choose them such that the two-mode squeezing process get into resonance. The Hamiltonian of Eq.~(\ref{eq1}) in the qubit approximation is rewritten as:
\begin{subequations}
\begin{eqnarray}
\label{eq3}
\hat{H}_{0}&=&\frac{\omega_{t}}{2}\hat{\sigma}_{z}^{\prime}+\omega_{1}\hat{b}_{1}^{\dagger}\hat{b}_{1}+\omega_{2}\hat{b}_{2}^{\dagger}\hat{b}_{2}, \\
\hat{H}_{1}&=&g_{01}(\hat{b}_{1}+\hat{b}_{1}^{\dagger})\hat{\sigma}_{z}^{\prime}+g_{02}(\hat{b}_{2}+\hat{b}_{2}^{\dagger})\hat{\sigma}_{z}^{\prime}, \\
\hat{H}_{d}&=&[E_{1}(t)+E_{2}(t)]e^{-i\omega_tt}\hat{\sigma}_{+}^{\prime}+\text{H.c.},
\end{eqnarray}
\end{subequations}
where the Pauli matrices $\{\hat{\sigma}_{x}^{\prime},\hat{\sigma}_{y}^{\prime},\hat{\sigma}_{z}^{\prime}\}$ have been introduced to identify the two charge isolated states of the transmon qubit, $\lbrace\vert0\rangle,\vert1\rangle\rbrace$.
We begin by moving into a proper rotating frame, which mathematically is achieved by applying the unitary transformation $\hat{U}= \exp\{\frac{i}{2}(\omega_{t}+\Delta)t\hat{\sigma}_{z}^{\prime}\}$ with $\Delta\equiv \omega_{1}-\omega_{2}$ to the whole Hamiltonian and we arrive at:
\begin{subequations}
\begin{eqnarray}
\label{eq4a}
\hat{H}_{0}&=&-\frac{\Delta}{2}\hat{\sigma}_{z}^{\prime}+\omega_{1}\hat{b}_{1}^{\dagger}\hat{b}_{1}+\omega_{2}\hat{b}_{2}^{\dagger}\hat{b}_{2}, \\
\hat{H}_{1}&=&g_{01}(\hat{b}_{1}+\hat{b}_{1}^{\dagger})\hat{\sigma}_{z}^{\prime}+g_{02}(\hat{b}_{2}+\hat{b}_{2}^{\dagger})\hat{\sigma}_{z}^{\prime}, \\
\hat{H}_{d}&=&[E_{1}(t)+E_{2}(t)]e^{i\Delta t}\hat{\sigma}_{+}^{\prime}+\text{H.c.}
\label{eq4c}
\end{eqnarray}
\end{subequations}
The above Hamiltonian can be diagonalized by choosing the following dressed basis:
\begin{subequations}
\begin{eqnarray}
\ket{e} &=& \cos\theta\ket{0} -\sin\theta\ket{1}, \\
\ket{g} &=& \sin\theta\ket{0} +\cos\theta\ket{1},
\end{eqnarray}
\end{subequations}
where $\theta = \frac{1}{2}\arctan\{\Lambda(t)/\Delta\}$ is the mixing angle with $\Lambda(t)=2\Big[\sum_{j,k}E_{j}(t)E_{k}^*(t)\Big]^{1/2}$. In this basis the effective frequency of the driven qubit is given by $\epsilon(t)=\sqrt{\Delta^2+\Lambda^{2}(t)}$ which is tunable by the external drive or the gate voltage.
Therefore, the driven system Hamiltonian in the dressed basis reads
\begin{align}
\label{eq6}
\tilde{H} =&-\frac{\epsilon(t)}{2}\hat{\sigma}_{z}+\omega_{1}\hat{b}_{1}^{\dagger}\hat{b}_{1}+\omega_{2}\hat{b}_{2}^{\dagger}\hat{b}_{2}
\nonumber\\
&-\hat\sigma_x\big[g_{1x}(\hat{b}_{1}+\hat{b}_{1}^{\dagger}) +g_{2x}(\hat{b}_{2}+\hat{b}_{2}^{\dagger})\big]
\end{align}
where
$g_{jx}(t)=g_{0j}\Lambda(t)/\epsilon(t)$ with $j=1,2$ and the Pauli matrices $\{\hat{\sigma}_{x},\hat{\sigma}_{y},\hat{\sigma}_{z}\}$ are defined over the dressed basis $\{\ket{g},\ket{e}\}$.
Note that to simplify the calculations we have dropped the terms describing the dispersive qubit--MR couplings in the above Hamiltonian as their sole effect is a shift in the frequency of the mechanical resonators in the resultant effective Hamiltonian.
The two effective coupling constants $g_{1x}(t)$ and $g_{2x}(t)$ as well as the effective spacing of the transmon qubit $\epsilon(t)$ can be well controlled by the gate voltage or the external drive.

We now apply the Fr{\"o}hlich-Nakajima approach to the Hamiltonian $\tilde{H}$. In fact, this takes us to the frame in which the first order interaction terms are zero and the remaining interaction terms in the transformed Hamiltonian are of the second and higher orders of $\upsilon_j \equiv g_{jx}/|\epsilon-\omega_j|$. Therefore, the transmon degree of freedom decouples from the mechanical resonators and the system can be evaluated in the Hilbert space related only to the mechanical modes, provided that $\upsilon_j \ll 1$.
In order to perform the transformation we write the Hamiltonian $\tilde{H}$ in Eq.~\eqref{eq6} as $\tilde{H}=\tilde{H}_{0}+\eta\tilde{H}_{1}$ where $\tilde{H}_0 = -[\epsilon(t)/2]\hat\sigma_z +\sum_j \omega_j \hat{b}_{j}^{\dag}\hat{b}_{j}$ and the second term describes the interaction between the transmon mode and each of the mechanical modes $\tilde{H}_{1} = \hat\sigma_x\sum_j g_{jx}(\hat{b}_j +\hat{b}_j^{\dag})$. 
Here, $\eta$ is introduced as a perturbation parameter to the interaction part of the Hamiltonian for controlling the order of expansions and shall be set to unity at the end of calculations.

The effective mechanical--mechanical Hamiltonian is obtained by applying the unitary transformation $\exp\{-\eta\hat{S}(t)\}$ to \eqref{eq6}.
The Hamiltonian in this rotating frame is given by
\begin{eqnarray}
\label{eq8}
\hat{H}'= e^{-\eta\hat{S}(t)}\tilde{H}e^{+\eta\hat{S}(t)} -i e^{-\eta\hat{S}(t)}\frac{\partial}{\partial t}\Big(\sum_{n=0}^{\infty}\frac{[\eta\hat{S}(t)]^{n}}{n!}\Big),
\end{eqnarray}
where the second term appears due to the time dependence of $\hat{S}$ operator.
By employing the Baker-Hausdorff formula and rearranging in orders of $\eta$ we arrive at
\begin{align}
\label{eq9}
\hat{H}' &= \tilde{H}_{0} +\eta(\tilde{H}_{1} +[\tilde{H}_{0},\hat{S}] -i\partial_{t}\hat{S}) \nonumber\\
		 &+ \eta^{2}\Big([\tilde{H}_{1},\hat{S}] +\frac{1}{2!}\big[\hat{S},[\hat{S},\tilde{H}_{0}]\big]\Big) +\mathcal{O}(\eta^{3}).
\end{align}
In the large detuning regime 
\begin{equation}
\label{eq9}
	\vert\epsilon(t)-\omega_{1}\vert\gg g_{1x}(t),~~\vert\epsilon(t)-\omega_{2}\vert\gg g_{2x}(t),
\end{equation}
where the Fr{\"o}hlich-Nakajima approach works well~\cite{Y.Li;2007} one can obtain the effective coupling between the two mechanical modes with applying the unitary transformation introduced above which the genarator $\hat{S}(t)$ is an anti-Hermitian operator.
This generator $\hat{S}(t)$ have to be chosen such that it satisfies the following equation
\begin{equation}
\label{eq10}
	\tilde{H}_{1}+[\tilde{H}_{0},\hat{S}]-i\hspace{0.5mm}\partial_{t}\hat{S}=0
\end{equation}
By this choice the transformed Hamiltonian $\hat{H}'$ up to the second order in $\eta$ reads
\begin{eqnarray}
\label{eq11}
\hat{H}'\approx\tilde{H}_{0}+\eta^{2}[\tilde{H}_{1},\hat{S}]+\frac{\eta^{2}}{2!}\big[\hat{S},[\hat{S},\tilde{H}_{0}]\big].
\end{eqnarray}
We further restrict our study to the case where $|\dot g_{jx}(t)/g_{jx}(t)|\ll|\epsilon(t)-\omega_{j}|$, which the time-dependent couplings change slowly~\cite{Y.Li;2007}. In this case we find the generator of the transformation $\hat{S}(t)$ from Eq.~(\ref{eq10}) as:
\begin{align}
\label{eq12}
\hat{S}(t) &\approx \sum_{j=1,2}g_{jx}(t)\bigg[\frac{\hat{b}_{j}^\dagger\hat{\sigma}_{+}-{\hat{b}}_{j}\hat{\sigma}_{-}}{\epsilon(t)+\omega_{j}}
+\frac{\hat{b}_{j}\hat{\sigma}_{+}-\hat{b}_{j}^\dagger\hat{\sigma}_{-}}{\epsilon(t)-\omega_{j}}\bigg].
\end{align}
In this regime, the driven qubit mostly remains in the dressed ground state $\vert g\rangle$. Hence, the effective Hamiltonian is attained as $\hat{H}'=\hat{H}_{\rm eff}\otimes \vert g\rangle\!\langle g\vert$.
In a frame rotating with the mechanical frequencies, the explicit form of the fully mechanical effective Hamiltonian reads:
\begin{align}
\label{eq13}
\hat{H}_{\rm eff} &=-G_1\hat{b}_{1}^{\dagger}\hat{b}_{1} +G_{2}\hat{b}_{2}^{\dagger}\hat{b}_{2}
\nonumber\\
&-\frac{1}{2}\Big(G_{1}\hat{b}_{1}^{2} e^{-2i\omega_{1}t} -G_{2}\hat{b}_{2}^{2} e^{-2i\omega_{2}t} +\text{H.c.}\Big)
\\
&-G_{12}\Big(\hat{b}_{1}\hat{b}_{2} e^{-i(\omega_{1}+\omega_{2})t}+\hat{b}_{1}^{\dagger}\hat{b}_{2} e^{i(\omega_{1}-\omega_{2})t} +\text{H.c.}\Big),
\nonumber
\end{align}
where we have introduced the frequency shift and squeezing factor $G_j(t) \equiv 2\epsilon(t) g_{jx}^{2}(t)/[\epsilon(t)^2 -\omega_{j}^2]$ and the effective mechanical-mechanical coupling strength
\begin{equation}
\label{eq14}
G_{12}(t)=g_{1x}(t)g_{2x}(t)\frac{\epsilon(t)(\omega_{1}^{2}-\omega_{2}^{2})}{[\epsilon(t)^{2}-\omega_{1}^{2}][\epsilon(t)^{2}-\omega_{2}^{2}]}.
\end{equation}
The first line in the above effective Hamiltonian shows a part shifting the frequency of each MR, which is identifiable in an experiment and can essentially be compensated for. The second line corresponds to the single-mode squeezing processes. The last line is composed of (i) a phonon hopping process, i.e. a beam-splitter Hamiltonian leading to the quantum tunneling of phonons between the two MRs, and (ii) a two-mode squeezing process, which can lead to the entanglement of the two MRs.
We note that the coupling rate given in Eq.~({\ref{eq14}) shows that the entangling process vanishes for equal mechanical frequencies $\omega_{1}=\omega_{2}$. 

A constant drive amplitude will only excite the phase shifting processes. However, to bring either of the single- and two-mode squeezing, or the beam-splitter processes into resonance one applies a modulated drive to the transmon mode with wisely chosen frequencies. In this paper, we are interested in generating a motional entangled state starting from initially separable state of the system. Therefore, it is necessary to only excite the two-mode squeezing process. For this purpose, we choose to parametrically drive the transmon at the sum of the two mechanical frequencies as it will become clear, shortly~\cite{Glave2010,Abdi et al.;2015,Tian2008}.

In fact, one needs to bring into resonance the following part of the effective Hamiltonian in Eq. (\ref{eq13}}):
\begin{equation}
	\hat{H}_{\rm TMS}=-G_{12}(t)\Big(\hat{b}_{1}\hat{b}_{2} e^{-i(\omega_{1}+\omega_{2})t} +\text{H.c.}\Big).
\label{eq15}	
\end{equation}		
We remind that the time dependence of $G_{12}(t)$ stems from $\Lambda(t)$, which in turn, depends on the transmon drives [see Eq. (\ref{eq14})]. In order to clarify our calculations, we simplify it by considering a harmonic oscillation with time for the two complex amplitudes as $E_{1}(t)=\mathcal{E}_1 e^{-i\omega_{L1}t}$ and $E_{2}(t)=\mathcal{E}_2 e^{i\omega_{L2}t}$. By this we are brought to
\begin{eqnarray}
\Lambda(t)&=&2\Big[\sum_{j,k}E_{j}(t)E_{k}^*(t)\Big]^{1/2}\hspace{4cm} \nonumber\\
	&=&2\sqrt{|\mathcal{E}_{1}|^{2}+|\mathcal{E}_{2}|^{2}+2\mathcal{E}_{1}\mathcal{E}_{2}\cos(\omega_{D}t)},
\label{eq16}
\end{eqnarray}
where $\omega_D\equiv\omega_{L1}+\omega_{L2}$.
By substituting from (\ref{eq16}) in (\ref{eq14}) it becomes clear that the two-mode squeezing process is excited only if:
\begin{equation}
	\omega_{D}=\omega_{1}+\omega_{2}.
\label{eq17}
\end{equation}
Therefore, one possible choice for the drive Hamiltonian in Eq.(\ref{eq4c}) that satisfies condition of Eq.~(\ref{eq17}) is to write it as the following:
\begin{equation}
	\hat{H}_{d}=[\mathcal{E}_{1}e^{-i\omega_{1}t}+\mathcal{E}_{2} e^{i\omega_{2}t}]e^{i\Delta t}\hat{\sigma}_{+}^{\prime}+\text{H.c.}
\label{eq18}
\end{equation}

By this choice the terms related to the TMS process are brought into resonance while the other terms are rapidly oscillating and thus have a negligible effect on the dynamics of system.

At this point, let us emphasize that the effective Hamiltonian \eqref{eq13} provides a general picture of the dynamics of the mechanical subsystem. The total system dynamics involves complexities of the transmon qubit and, because of coupling to the environments, experiences irreversible and phase destroying dynamics. For this reason, in Sec.~\ref{nr} the full system is numerically simulated.

\section{Measures OF NON-GAUSSIANITY and entanglement \label{measures}}
\subsection{Non-Gaussianity \label{nG}}
Generally, dynamics of systems whose Hamiltonian can not be written as second-order terms or lower in quadrature operators of the system is nonlinear. Therefore, one expects that their initial Gaussian state evolve into non-Gaussian states at later times.
On this basis many state generation protocols designed for preparation of non-Gaussian states employ nonlinear interactions. That is, those with interaction Hamiltonian of higher orders than two in quadrature operators~\cite{Gerrits,Sasaki} such as Kerr effect~\cite{GEnoni,Allevi,Deleglise} or trilinear light-matter interaction induced by radiation pressure in optomechanical systems~\cite{Mancini,Bose,Ludwig}.
In our system, due to the nonlinear nature of the transmon qubit as well as its nonlinear interaction with the harmonic mechanical oscillators one anticipates that the state of fully mechanical bipartite subsystem evolve into a non-Gaussian state. Non-Gaussianity of a state can be measured by identifying its distance from the closest Gaussian state with similar properties, the reference state. Depending on the property that is taken to calculate this distance various measure can be defined. In this section, we briefly review the approach which will be used for measuring the non-Gaussianity (NG) of the mechanical states under study based on von Neumann entropy~\cite{Genoni;2010}.

The degree of NG of a state can be quantified by measuring its relative entropy with respect to the reference Gaussian state. In fact, one quantifies the NG of a quantum state $\rho$ in terms of its entropy difference with respect to a reference Gaussian state $\rho_{G}$, whose first and second moments are the same as those of the original state $\rho$:
\begin{equation}
	\delta=S(\rho_{G})-S(\rho),
\label{eq19}
\end{equation}
where $S(\rho)=-\tr\big\{\rho\log(\rho)\big\}$ is the von Neumann entropy of $\rho$.
In our system the NG is computed by first numerically solving the master equation in Eq.~(\ref{eq2}) and obtaining the mechanical reduced density matrix $\rho_{12}(t)$ by tracing out the transmon degrees of freedom. We then construct the corresponding reference Gaussian state $\rho_G(t)$ by computing the first and second moments of the mechanical operators at the state $\rho_{12}(t)$ to attain the covariance matrix (CM) $\bm\sigma$ at every instance of time. Elements of the CM are given by
\begin{equation}
\sigma_{ij}(t)=\langle\lbrace\hat{X}_i,\hat{X}_j^\dag\rbrace\rangle_{\rho_{_{12}}} -2 \langle\hat{X}_i\rangle\langle\hat{X}_j^\dag\rangle_{\rho_{_{12}}},
\end{equation}
where $\lbrace\cdot,\cdot\rbrace$ is the anti-commutator while $\boldsymbol{\hat{X}}=(\hat{b}_{1}, \hat{b}_{2},\hat{b}_{1}^{\dagger},\hat{b}_{2}^{\dagger})^{\intercal}$ is the vector of system operators.
The von Neumann entropy of a Gaussian state is easily calculated by finding the symplectic eigenvalues of the CM that are of the form $\lbrace\pm\nu_{+},\pm\nu_{-}\rbrace$ and putting them into~\cite{Genoni;2010,Adesso,Serafini}
\begin{equation}
S(\rho_{G})= h(\nu_{+})+h(\nu_{-}),
\label{eq20}
\end{equation} 
where $h(x)=\frac{x+1}{2}\log(\frac{x+1}{2})-\frac{x-1}{2}\log(\frac{x-1}{2})$.
These eigenvalues fulfill the condition $\nu_\pm\geq 1$ for all physical states~\cite{Adesso}. In our chosen basis $\bm{\hat{X}}$, the symplectic eigenvalues are indeed the eigenvalues of $i\boldsymbol{\Omega}\boldsymbol{\sigma}(t)$ where $\boldsymbol{\Omega}=\text{diag}(-i,-i,i,i)$ is the symplectic form.
By substitution in Eq.~(\ref{eq19}) the NG of the mechanical state can be measured by $\delta_{12}(t)=S(\rho_{G}(t))-S(\rho_{12}(t))$.
The state $\rho_{12}(t)$ is non-Gaussian if and only if $\delta_{12}(t)>0$.

\subsection{Entanglement}
\label{EN}
In order to quantify the degree of entanglement between the two MRs, we opt to use logarithmic negativity for the reduced density matrix of the fully mechanical bipartite subsystem~\cite{Vidal et al.;2002, Plenio2005}:
\begin{equation}
\label{logneg}
E_{N}(\rho_{12})=\log_{2}\lVert\rho_{_{12}}^{\intercal_{_2}}\rVert_{1},
\end{equation}
where $\rho_{_{12}}^{\intercal_{_2}}$ is the partial transpose of the reduced density matrix with respect to one of its subsystems (here the second mechanical resonator) and $\lVert\cdot\rVert_{1}$ denotes the trace norm.

We numerically calculate the logarithmic negativity for the reduced density matrix of the fully mechanical bipartite subsystem by first numerically solving the Lindblad master equation introduced in Sec.~\ref{model} and then partial tracing on the transmon mode. The outcome is a reduced density matrix $\rho_{12}$ describing state of the mechanical bipartite subsystem. The amount of entanglement between the MRs is measured by setting the state in Eq.~(\ref{logneg}).

\section{NUMERICAL RESULTS\label{nr}}
Here, we numerically verify ability of the protocol proposed in section~\ref{fmec} for generating entangled mechanical states and survey properties of the final state of the micromechanical resonators. To this end, we solve the Lindblad master equation~(\ref{eq2}) with the original Hamiltonian of the whole system given in Eq.~(\ref{eq1}). The simulations are performed using the quantum toolbox in Python (QuTiP)~\cite{Johansson;2013}.
To start with, we assume that the MRs are initialized closed to their ground-state by a sideband cooling technique~\cite{Abdi et al.PRL;2015} and the transmon qubit is initialized in its ground state. Therefore, the initial state of the system is set to $\vert\psi_{0}\rangle=\vert0\rangle_{t}\vert0\rangle_{1}\vert0\rangle_{2}$, which obviously is a separable state.
%Here, $\vert0\rangle_{t}$ means that the transmon qubit is in the charge-isolated initial state.
%figure2-------------------------------------------------------
\begin{figure}[b!]
\includegraphics[width=0.9\columnwidth]{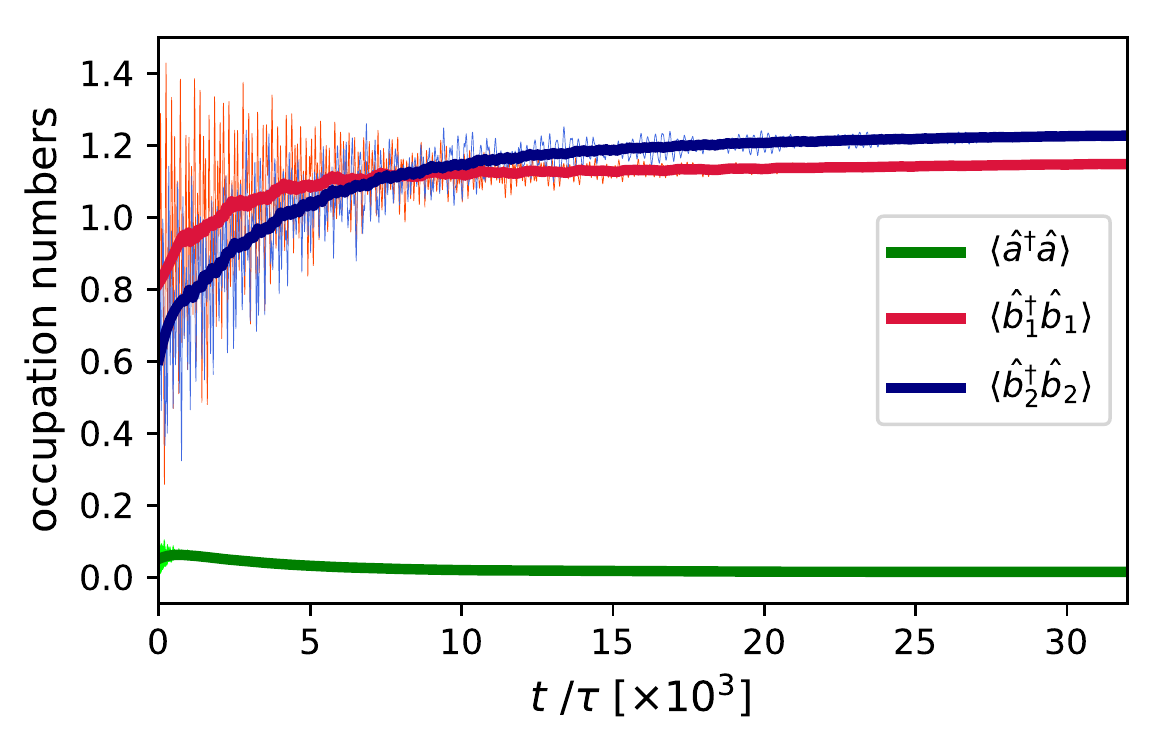}
\caption{%
Stability of the system: Dynamics of the expectation values of the bosonic number operators for the transmon qubit (green), MR1 (red), and MR2 (blue) for the parameters given in the text with $\Delta g= 13.9$~kHz. After a long time, the system observables tend to a quasi-stationary value. The time has been normalized to $\tau=\frac{2\pi}{\omega_{1}+\omega_{2}}$.}
\label{fig2}
\end{figure}
%------------------------------------------

The system parameters are chosen such that the system remains far from any instability. The exact determination of the stability regions for such nonlinear systems is a tedious task. Nonetheless, one applies a mean-field approximation to linearize the system dynamics. Then the stability regions are straightforward to be identified by the Routh-Hurwitz criterion~\cite{Dejesus}.
Furthermore, after identifying the approximate regions of stability for the parameters of our system, we numerically check and verify dynamical stability of the system by tracking time evolution of the expectation values for a few physical observables of our system. For a quasi-stable system, the observables are bound to converge to a finite value. The tracking is done for every simulation reported in this manuscript.

In what follows, we choose experimentally feasible values of parameters; The mechanical mode frequencies are taken to be slightly different to ensure transmon-mediated mechanical--mechanical coupling (see the above discussion) $\omega_{1}/2\pi = 10~\text{MHz},~\omega_{2}/2\pi= 9.95~\text{MHz}$ with a quality factor $Q=2\times 10^5$. The thermal bath occupation number are set to $\bar{n}_{1}\approx\bar{n}_{2}\equiv n_{\rm th}\approx 0.2$ phonons for the MRs.
The effect of higher mechanical thermal noise will be discussed at the end of this section.
%This corresponds to an effective bath temperature of $T=0.1$~mK due to the passive cooling of the mechanical resonators. This can be done, e.g. by an extra drive tone or an auxiliary qubit~\cite{Abdi2017}.
The Josephson and charge energies of the transmon qubit are such that $\omega_{t}/2\pi=17$~GHz and $\lambda/2\pi = 0.25~\text{GHz}$ similar to the values in the Ref.~\cite{Abdi et al.PRL;2015}.
The qubit is subject to an environmentally induced relaxation rate $\gamma_{t}/2\pi= 4.5~\text{kHz}$ and pure dephasing $\gamma_{\phi}=2\gamma_{t}$. The qubit is bi-chromatically driven with real amplitudes $\mathcal{E}_{1}\approx\mathcal{E}_{2} = 8.0~\text{MHz}$.
The strength of coupling between the qubit and MRs is first equally set to $g_{01}=g_{02}=325.6~\text{kHz}$. However, we introduce asymmetries in the coupling to study its effect on the mechanical--mechanical entanglement dynamics.
%%----------figure3----------------------------------------------
\begin{figure}[t!]
\includegraphics[width=0.9\columnwidth]{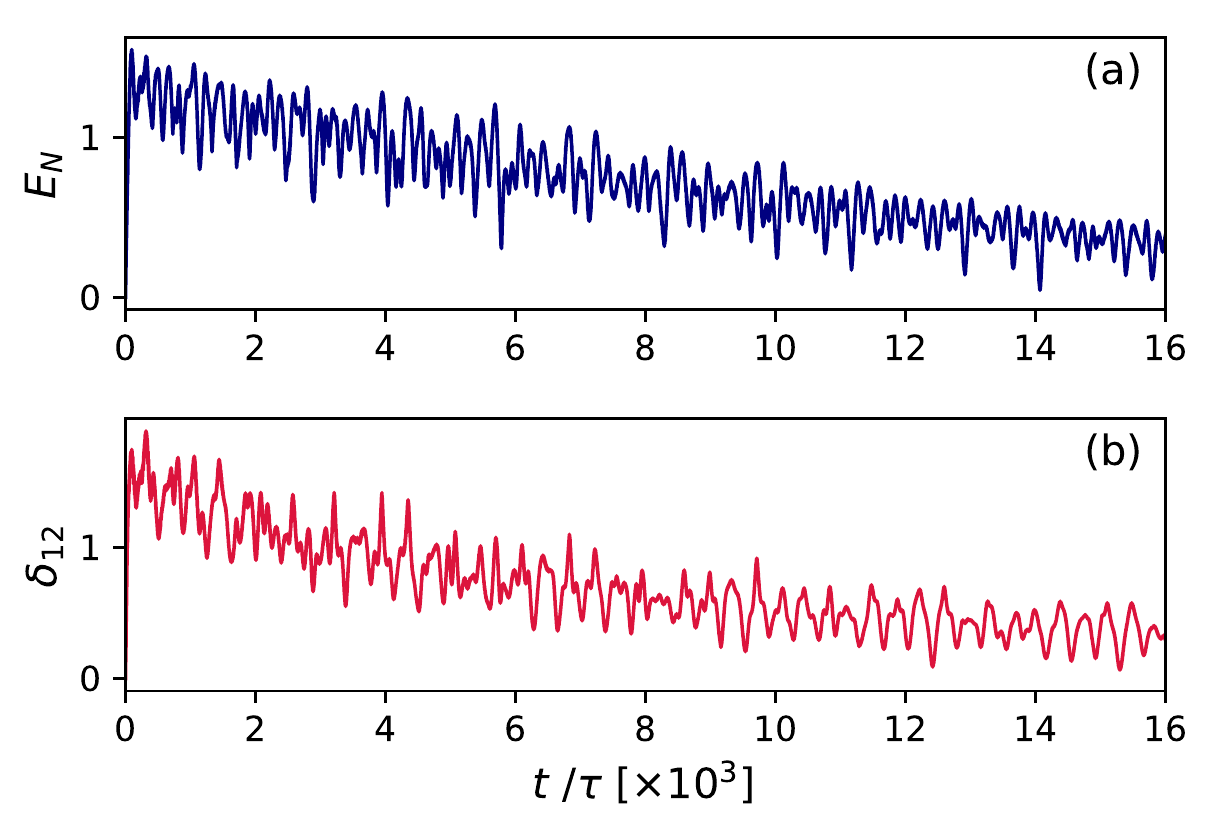}
\caption{%
Dynamics of (a) the entanglement of motion between the mechanical resonators measured by logarithmic negativity $E_{N}$ and (b) the NG quantified by the measure $\delta_{12}$ for the parameters given in the text.
}
\label{fig3}
\end{figure}
%--------------------------------------------------------------
As mentioned above to check of the system stability in the considered parameter region, we numerically evaluate the expectation values of the transmon occupation number $\mean{\hat{a}^{\dag}\hat{a}}$ and the bosonic number operators of the mechanical modes $\mean{\hat{b}_1^{\dag}\hat{b}_1}$ and $\mean{\hat{b}_2^{\dag}\hat{b}_2}$ over the long times.
Figure~\ref{fig2} represents an example of our calculations to verify the dynamical stability of our system for $\Delta g =|g_{01}-g_{02}|=13.9~\text{kHz}$ and the other parameters same as those introduced above. This plot demonstrates that the time evolution of the occupation operators converge into a finite value, starting from the separable initial state. For the rest of parameters the same examination procedure has been performed and stability of the system has been ensured. It should be noted that in all of the plots presented in this paper, the time has been normalized to $\tau=2\pi/(\omega_{1}+\omega_{2})$, which equals period of the mechanical TMS process.

The various numerical results show that the stability of the whole system does not change for our chosen drive amplitude with changing the mechanical frequencies up to a few MHz, the energy relaxation and dephasing rate of the transmon qubit and the damping rates of the MRs up to one order of magnitude larger or smaller than the values given above. Also the values of the coupling rates $g_{01}$ and $g_{02}$ up to about a few MHz.
The large stability region of the system stability further proves its experimental feasibility and flexibility.
 
In Fig.~\ref{fig3}(a) dynamics of the mechanical-mechanical entanglement for the parameters given above is presented. Also according to the approach introduced in Sec.~\ref{measures} to quantify the NG of the state of the subsystem composed of two micromechanical resonators, we numerically obtain the reduced density matrix $\rho_{12}(t)$ at every instance of time $t$ from the solution of the master equation (\ref{eq2}) with the initial state $\vert\psi_{0}\rangle$.
We then calculate its second moments to construct the covariance matrix and by computing its symplectic eigenvalues the Von Neumann entropy of the reference Gaussian state ($\rho_{G}(t)$) is obtained. The amount of the NG of the state $\rho_{12}(t)$ is eventually calculated from Eq.~(\ref{eq19}) at each step of time. Time evolution of the non-Gaussianity is illustrated in Fig.~\ref{fig3}(b).
%%%%----------figure4--------------------------------------------
\begin{figure}[b!]
\includegraphics[angle=0, width=0.9\columnwidth]{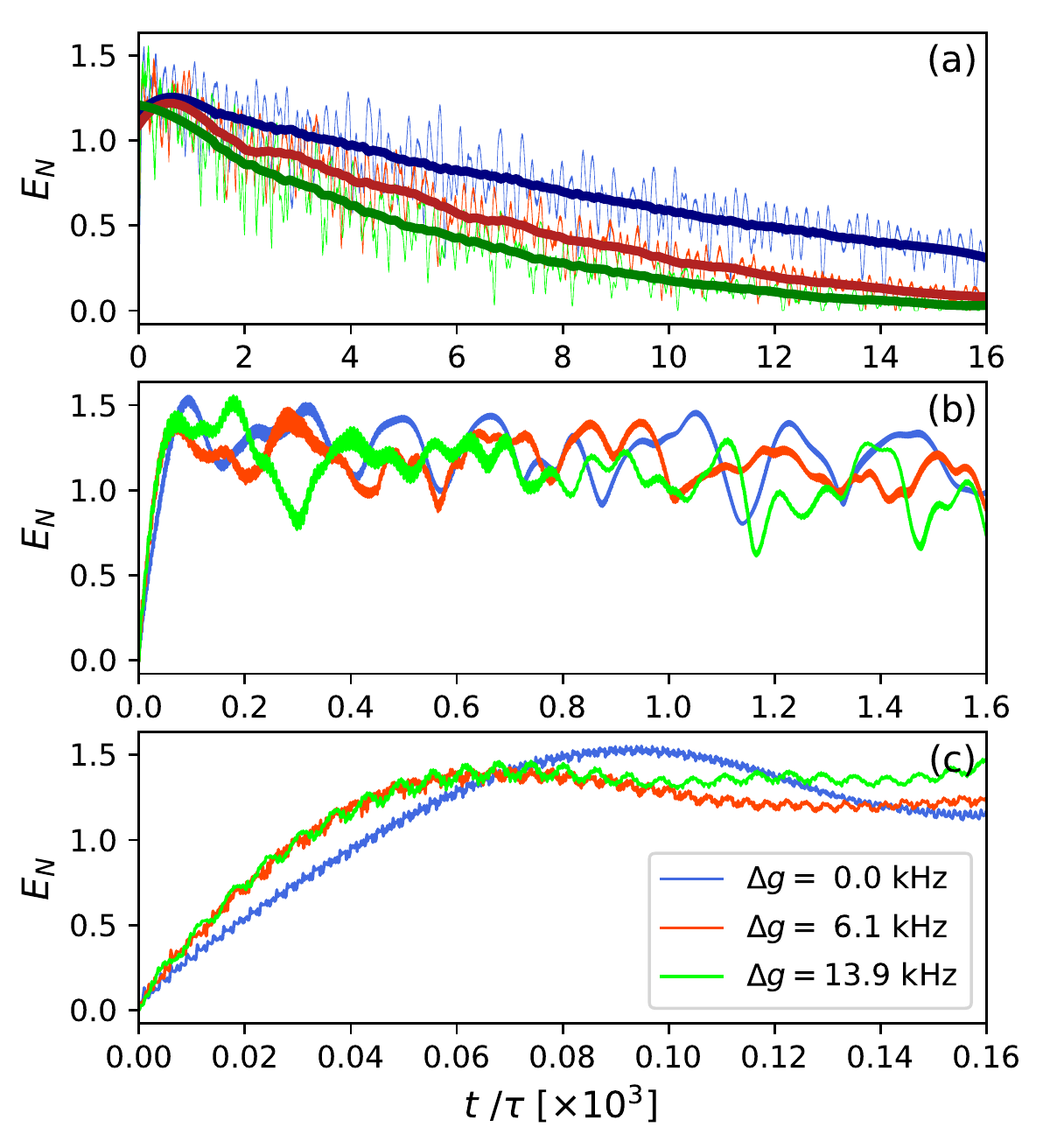}
\caption{%
Time evolution of the mechanical logarithmic negativity $E_{N}$ for three different values of $\Delta g= \big(0.0,~6.1,~\text{and}~13.9\big)$~kHz at long (a), intermediate (b), and short (c) time periods. The solid lines in (a) show the trends of $E_N$.
The rest of parameters are the same as those given in the text.
}
\label{fig4}
\end{figure}
%%%%------------------------------------------------------------
Figure~\ref{fig3} verifies creation of a large non-Gaussian entanglement between the two mechanical resonators, starting from a Gaussian separable state. Indeed $|\psi_0\rangle$ is converted into an inseparable, non-Gaussian state thanks to the nonlinear Hamiltonian Eq.~(\ref{eq1}) and properly driven qubit.
The lifetime of the entangled state is limited by the qubit coherence time and for the relaxation rate we are considering here can last for a few tens of milliseconds.
We notice that $\delta_{12}$ follows a pattern nearly similar to that of logarithmic-negativity $E_{N}$. However, tending towards a nonzero asymptotic quasi-stationary value. Such that the final motional state remains non-Gaussian, though separable.

We next examine effect of imperfections on the amount of dynamics of the entanglement.
In experimental implementation of our setup one of the most probable imperfection is having the mechanical resonators asymmetrically coupled to the transmon qubit.
Indeed, our results show that the maximum entanglement is achieved for a symmetric system. That is, for $\Delta g \equiv |g_{01} - g_{02}| =0 $. More interestingly, the resultant state exhibits a longer lifetime.
In order to study the effect of asymmetry $\Delta g$, we plot time evolution of the logarithmic negativity of $\rho_{12}(t)$ for three different values of the coupling rate deviation $\Delta g$, while the other parameters are kept the same as those given above.
In Fig.~\ref{fig4} the results are summarized where the time evolution of the mechanical logarithmic negativity is plotted for the three values $\Delta g=0.0,~6.1$, and $13.9$~kHz.
The $E_N$ is plotted at three different time scales: long, intermediate, and short.
Even though the entanglement dynamics are not different at short and intermediate time periods for the three different $\Delta g$ values, a drastic collapse in the entanglement lifetime is clear from the plots as the coupling rates set to differ $\Delta g \neq 0$.
Therefore, when it comes to the long-living entangled states smaller differences in the coupling rates are more favorable.
%%----------figure5----------------------------------------------
\begin{figure}[t!]
\includegraphics[width=0.9\columnwidth]{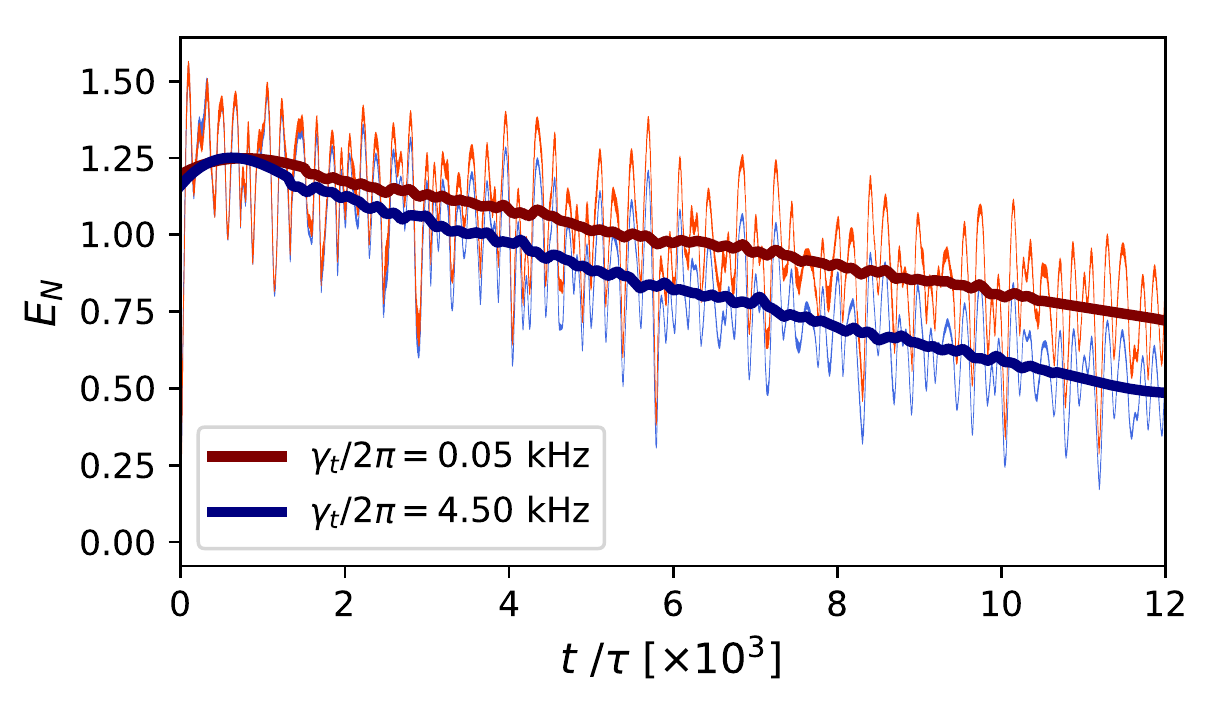}
\caption{%
Dynamics of the entanglement measured by $E_{N}$ for two different qubits: one with $\gamma_{t}/2\pi=0.05$~kHz (red line) and $\gamma_{t}/2\pi=4.5$~kHz (blue line). The pure dephasing rate in both cases is set to $\gamma_\phi=2\gamma_t$ and other parameters are the same as the parameters of the text.
}
\label{fig5}
\end{figure}
%-------------------------	

As another limiting factor, we investigate the effect of transmon decoherence time on the entanglement. The simulation outcomes show that by decreasing relaxation rate and dephasing rate of the transmon qubit the mechanical--mechanical entanglement lives longer. In Fig.~\ref{fig5} dynamics of $E_N$ is plotted for two different superconducting transmon qubits: one with $\gamma_{t}/2\pi=4.5$~kHz and $\gamma_\phi/2\pi=2.3$kHz, and an optimistically chosen values of $\gamma_t/2\pi=0.05$~kHz and $\gamma_\phi/2\pi=0.03$~kHz, which is within reach.	

\begin{figure}[t!]
\includegraphics[width=0.9\columnwidth]{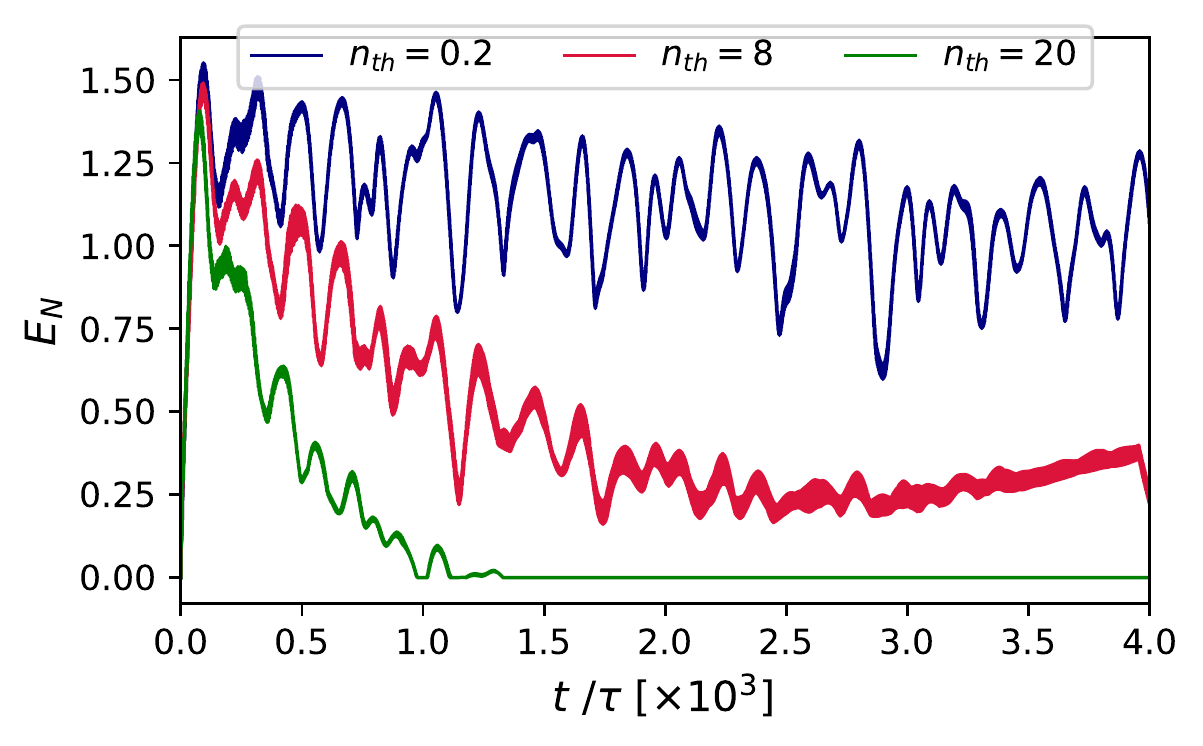}
\caption{%
Variations of the mechanical--mechanical entanglement dynamics with thermal noise: $E_N$ is plotted for three different values of mechanical thermal occupation numbers $n_{\rm th}=0.2$ (blue), $n_{\rm th}=8$ (red), and $n_{\rm th}=20$ (green). The last one corresponds to a bath temperature of $T\approx10$~mK, which is experimentally available. The other parameters are the same as those given in the text.
}
\label{fig6}
\end{figure}

The mechanical damping is not expected to considerably affect the general system dynamics in the course of qubit decoherence time, and the numerical results verify this statement.
Nonetheless, when it comes to the entanglement, thermal noise is one of the most prohibiting environmental effects. We, thus, study the dimensions of its effect on the entanglement dynamics numerically. As stated above, we assume that the mechanical resonators are cooled-down very close to their ground state, a task that is doable by side-band technique~\cite{Rabl2010, Abdi et al.PRL;2015}. 
We further assumed that during excitation of TMS for generation of the entanglement a cooling procedure is still working.
In order to study the effect of the thermal noise on the entanglement dynamics without permanently cooling mechanical resonators, we consider two other thermal environment temperatures and evaluate the logarithmic negativity for the mechanical subsystem.
For the sake of simplicity of comparison we take the same parameter values used in Fig.~\ref{fig3} but with three different phonon thermal occupation numbers $n_{\rm th}=0.2,~8$, and $20$. These values correspond to the environment temperatures of $T\approx0.1,~3.8$, and $10$~mK, respectively, which the last one is experimentally available.
Fig.~\ref{fig6} shows the results. Comparing the plots, one clearly observes dramatic dependence of $E_N$ on the thermal noise.
Interestingly, despite fragility of the mechanical-mechanical entanglement at a bath temperature of $T=10$~mK, it rises to relatively high values at short times before collapsing into a separable state at a time scale of about $300$~microseconds.
%%----------figure6----------------------------------------------
%-------------------------

\section{CONCLUSION \label{conclusion}}
In summary, we have proposed a protocol to generate a \textit{motional non-Gaussian entangled state} of two \textit{massive micromechanical resonators} in an electromechanical hybrid device.
The entanglement is achieved by interposing a superconducting qubit between two micromechanical resonators.
In our protocol, the intermediate transmon qubit is driven by two time-modulated microwave pulses whose frequencies are devised such that a two-mode squeezing process between the two MRs is excited.
We have numerically verified the performance of our scheme.
The results of simulations demonstrate the possibility of generation of an appreciable non-Gaussian mechanical--mechanical entanglement in a wide range of parameters without entering the instability region of the system.
The entanglement is achieved for experimentally feasible parameters, and more interestingly, the quasi-stationary entangled state is non-Gaussian, making it an appropriate resource for universal CV quantum information processing.
Dependence of the amount and lifetime of the entanglement on the mechanical coupling rate asymmetry, the transmon qubit decoherence rate, and the mechanical thermal noise has been studied.
Despite reduction in the quantity and quality of the entanglement in the presence of imperfections, the scheme shows appreciable entanglement and non-Gaussianity properties.

\begin{acknowledgments}
The work of NES and MKN has been supported by a research grant from Ferdowsi University of Mashhad.
MA acknowledges support by STDPO and IUT through SBNHPCC.
\end{acknowledgments}

\appendix*
\section*{Appendix: Derivation Of the system Hamiltonian}
\label{App1}
We consider the hybrid electromechanical system depicted in Fig. (\ref{fig1}) and would like to obtain the Hamiltonian of Eq.(\ref{eq1}) in the main text. The Hamiltonian of the system composed of a transmon qubit which has two shunt capacitor whose shunt capacitances depend on the position of the mechanical resonators is given by
\begin{eqnarray}
\label{eqA.23}
\hat{H}^{\prime}_{0}&=&4E_{C}(\hat{x}_{1},\hat{x}_{2})(\hat{n}-n_{g})^{2}-E_{J}\cos\hat{\varphi} +\hbar\omega_{1}\hat{b}_{1}^{\dagger}\hat{b}_{1}+\hbar\omega_{2}\hat{b}_{2}^{\dagger}\hat{b}_{2}
	\nonumber\\
	&&+\big[E_{1}(t)+E_{2}(t)\big](\hat{a}+\hat{a}^{\dagger}),
\end{eqnarray}
where the last line is related to the coherent drive Hamiltonian applied to the transmon mode. The superconducting charge number and phase operators are denoted by $\hat{n}$ and $\hat{\varphi}$, respectively, which satisfy the commutation relation $[\hat{\varphi},\hat{n}]=i$ and can be defined in terms of annihilation (creation) operators $\hat{a}$ ($\hat{a}^{\dagger}$) of the transmon mode in the following way:
\begin{eqnarray}
\label{eqA.24}
	\hat{n}=\frac{1}{2}(\frac{E_{J}}{2E_{C}})^{1/4}(\hat{a}+\hat{a^{\dagger}})
	\nonumber\\
	\hat{\varphi}=i(\frac{2E_{C}}{E_{J}})^{1/4}(\hat{a}-\hat{a}^{\dagger})
\end{eqnarray}
In Eq.~\eqref{eqA.23} $n_{g}$ is the offset charge of the device which contains both dc and ac parts. Since in our protocol the transmon mode is driven by the classical bichromatic drive the resultant offset charge is classical. Applying the unitary gauge transformation $\hat{U}_G=e^{-in_{g}\hat{\varphi}}$, one can remove the offset charge $n_{g}$ in the above Hamiltonian~\cite{Girvin}:
\begin{eqnarray*}
		\hat{U}_G (\hat{n}-n_{g})^{2} \hat{U}_G^{\dagger}=\hat{n}^{2}.
\end{eqnarray*}

Moreover, since our system is operated in the transmon regime ($\zeta\equiv\frac{E_{J}}{E_{C}}\gg1$), we can expand the cosine term in the above Hamiltonian around $\varphi=0$ and keep up to the fourth order in $\hat\varphi$. 
	
One can also Taylor expand the charging Energy $E_{C}$ up to the first power in $\hat{x}_{1}$ and $\hat{x}_{2}$ as:
\begin{eqnarray*}
	E_{C}(\hat{x}_{1},\hat{x}_{2})\approx E_{C}+g^{\prime}_{01}	(\hat{b}_{1}+{\hat{b}_{1}}^{\dagger})+	g^{\prime}_{02}	(\hat{b}_{2}+{\hat{b}_{2}}^{\dagger})
\end{eqnarray*}
where $g^{\prime}_{0j}=E_{C} \frac{C_{Bj}}{C_{\Sigma}} \frac{x_{{\rm zp},j}}{d_{0j}}$ ($d_{0j}$ is the equilibrium distance between the plates of the shunt capacitor with the capacitance $C_{Bj}$). Substituting the above cases in Eq.~\eqref{eqA.23}, the Hamiltonian is given as the following:
\begin{eqnarray}
\label{eqA.25}
	\hat{H}^{\prime}_{1}&=&4E_{C}\hat{n}^{2}+E_{J}(1+\frac{\hat{\varphi}^{2}}{2!}-\frac{\hat{\varphi}^{4}}{4!})\nonumber\\
	&&+\hbar\sum_{j=1,2}\omega_{j}\hat{b}_{j}^{\dagger}\hat{b}_{j}+4\sum_{j=1,2} g^{\prime}_{0j}(\hat{b}_{j}+\hat{b}_{j}^{\dagger})\hat{n}^{2}\nonumber\\
	&&+\big[E_{1}(t)+E_{2}(t)\big](\hat{a}+\hat{a}^{\dagger}).
\end{eqnarray}		
Using (\ref{eqA.24}) we rewrite the above Hamiltonian in terms of the transmon bosonic annihilation and creation operators up to a constant as
\begin{align}
\label{eqA.26}
\hat{H}^{\prime}_{2}=&\sqrt{\frac{E_{J}E_{C}}{2}}(\hat{a}+\hat{a}^{\dagger})^{2}-\sqrt{\frac{E_{J}E_{C}}{2}}(\hat{a}-\hat{a}^{\dagger})^{2}-\frac{E_{C}}{12}(\hat{a}-\hat{a}^{\dagger})^{4}\nonumber\\
&+\hbar\sum_{j=1,2}\omega_{j}\hat{b}_{j}^{\dagger}\hat{b}_{j}+\sqrt{\frac{E_{J}}{2E_{C}}}\sum_{j=1,2}g^{\prime}_{0j}(\hat{b}_{j}+\hat{b}_{j}^{\dagger})(\hat{a}+\hat{a}^{\dagger})^{2}\nonumber\\
&+\big[E_{1}(t)+E_{2}(t)\big](\hat{a}+\hat{a}^{\dagger}),
\end{align}	
which after applying the RWA---valid in our working regime in which $E_{J}\gg E_{C}$ and $g_{01},g_{02}\ll \omega_{t}$.
By introducing $\hbar g_{0j}\equiv g^{\prime}_{0j}\sqrt{2 \zeta}$, $\hbar \lambda=E_{C}/2$, and $\hbar\omega_{t}=(\sqrt{8\zeta}-1)E_{C}$, the Hamiltonian is simplified to
\begin{align}
\label{eqA.27}
\hat{H}^{\prime}_{3}&=\hbar\omega_{t}\hat{a}^{\dagger}\hat{a}-\hbar\lambda \hat{a}^{\dagger 2}\hat{a}^{2}
	+\hbar\!\sum_{j=1,2} \omega_{j}\hat{b}_{j}^{\dagger}\hat{b}_{j}\nonumber\\
&+\hbar\!\sum_{j=1,2}g_{0j}\hat{a}^{\dagger}\hat{a}(\hat{b}_{j}+\hat{b}_{j}^{\dagger})+\hbar\sum_{j=1,2}g_{0j}(\hat{b}_{j}+\hat{b}_{j}^{\dagger})\nonumber\\
&+\big[E_{1}(t)+E_{2}(t)\big](\hat{a}+\hat{a}^{\dagger}).
\end{align}	
Thus, with neglecting the terms $\hbar\!\sum_{j}g_{0j}(\hat{b}_{j}+\hat{b}_{j}^{\dagger})$ whose sole effect is a slight modification in the position of the MRs the Hamiltonian of Eq.~(\ref{eqA.27}) reduces to that of Eq.~(\ref{eq1}) in the main text.

\end{document}